\documentclass[preprint,           
               noshowpacs,          
               preprintnumbers,     
               aps,                 
               prd,                 
               a4paper,             
               superscriptaddress,  
               nofootinbib,         
               tightenlines,        
               floats,floatfix      
               ]{revtex4}

\usepackage{graphics}
\usepackage[dvips]{graphicx}
\usepackage{latexsym}
\usepackage{amsmath,amssymb}        
\usepackage[draft=false]{hyperref}

\def\be{\begin{equation}}
\def\ee{\end{equation}}
\def\bea{\begin{eqnarray}}
\def\eea{\end{eqnarray}}
\newcommand{\ud}{\mathrm{d}}

\newcommand{\nn}{\nonumber \\}

\begin{document}

\title{Cosmological Constraints on $f(G)$ Dark Energy Models}

\author{Shuang-Yong Zhou\footnote{ppxsyz@nottingham.ac.uk}, Edmund J.~Copeland\footnote{ed.copeland@nottingham.ac.uk}, Paul M.~Saffin\footnote{paul.saffin@nottingham.ac.uk}}
\affiliation{School of Physics and Astronomy, University of Nottingham, University Park, Nottingham, NG7 2RD, UK}


\begin{abstract}

Modified gravity theories with the Gauss-Bonnet term $G=R^2-4R^{\mu\nu}R_{\mu\nu}+R^{\mu\nu\rho\sigma}R_{\mu\nu\rho\sigma}$ have recently gained a lot of attention as a possible explanation of dark energy. We perform a thorough phase space analysis on the so-called $f(G)$ models, where $f(G)$ is some general function of the Gauss-Bonnet term, and derive conditions for the cosmological viability of $f(G)$ dark energy models. Following the $f(R)$ case, we show that these conditions can be nicely presented as geometrical constraints on the derivatives of $f(G)$. We find that for general $f(G)$ models there are two kinds of stable accelerated solutions, a de Sitter solution and a phantom-like solution. They co-exist with each other and which solution the universe evolves to depends on the initial conditions. Finally, several toy models of $f(G)$ dark energy are explored. Cosmologically viable trajectories that mimic the $\Lambda$CDM model in the radiation and matter dominated periods, but have distinctive signatures at late times, are obtained.

\end{abstract}

\maketitle


\section{Introduction}

The discovery that our universe recently started accelerating \cite{Efstathiou:1990xe,snae,sdss,wmap5} has baffled particle physicists and cosmologists for a decade. Many suggestions about the origin of the cosmic acceleration, or the nature of dark energy, have been proposed (for a review, see \cite{dy123}). While the cosmological constant ($\Lambda$CDM) introduced by Einstein is the simplest proposal, albeit requiring fine-tunings to avoid coincidence problems \cite{finetuning}, other exotic negative-pressure fluids, often described in terms of scalar fields \cite{dydark}, have been proposed to address these issues. However, it is probably fair to say that no compelling and well-motivated solutions have yet been developed. As a result, attention has naturally turned to the possibility of dark energy originating from modifications of gravity itself \cite{DDG,cdtt,Capo2,fl02,ADDG,cfdett}.

Among the recent modifications of general relativity, perhaps the simplest approach is to replace the Einstein-Hilbert Lagrangian density with some general function of the Ricci scaler $\sqrt{-g}f(R)$ \cite{cdtt,Capo2,Capo1,nood2,Carloni:2004kp,hs07,ab07,staro07}. For these kind of models, generally it is easy to obtain a late time accelerated epoch. In \cite{fofr,Carloni:2007br}, conditions for the cosmological viability of general $f(R)$ models have been established. But as with the popular model $f(R)=R+\mu/ R$, they are also generally found to be inconsistent with solar system constraints \cite{solar1,solar2}.

A well motivated curvature invariant beyond the Ricci scalar is the the Gauss-Bonnet term
\be
G=R^2-4R^{\mu\nu}R_{\mu\nu}+R^{\mu\nu\rho\sigma}R_{\mu\nu\rho\sigma}\,,
\ee
which is inspired by string theory \cite{mettsey,Nojiri:2006je} and has gained special interest in cosmology \cite{carcopemad,anrita,tsu02,nood4,nood5,nood6}. It is also well known that the Gauss-Bonnet combination can reduce the number of spin-2 ghosts, which otherwise severely haunt the perturbation theory \cite{ghostgb}. The 4 dimensional Gauss-Bonnet term is a topological invariant, thus has no dynamical effects if added into the Lagrangian linearly. To introduce extra dynamics, one may couple the Gauss-Bonnet term to a scalar field, as naturally appears in low-energy string effective actions \cite{mettsey}. For the exponential coupling with a scalar field potential, this model can produce a matter dominated period followed by an accelerated period \cite{km07,san07}. However, it has been shown that it tends to conflict with solar system constraints \cite{acd12}.

Alternatively, we may consider the Lagrangian density as some general function of the Ricci scalar and the Gauss-Bonnet term $\sqrt{-g}F(R,G)$ \cite{cfdett,nood6}, and as a natural simplification, $F(R,G)=R/2+f(G)$ has been investigated \cite{nood5,dfhin07,libamo,davis0709,Nojiri:2007bt}. Uddin {\it et al.} \cite{lidsey08} studied the existence and stability of power-law scaling solutions in $f(G)$ models, finding that scaling solutions exist in the model $f(G)=\pm 2\sqrt{\alpha G}$ with $\alpha$ an arbitrary constant. In \cite{deftsu08}, the authors derived the stability conditions for a de Sitter solution and a standard matter/radiation solution in general $f(G)$ models; the authors also constructed several cosmologically viable $f(G)$ models, but found it difficult to simulate the models to high redshifts.

In this paper, we systematically survey the cosmological viability of general $f(G)$ models. The layout of this paper is as follows. In Section \ref{dframe}, we first derive field equations of general $f(G)$ models in the flat Robertson-Walker background and recast them nicely into a 5 dimensional autonomous system. Analogous to the case of $f(R)$ models \cite{fofr}, we construct the curve $m(r)$, where $m=Gf_{GG}/f_G$ and $r=-Gf_G/f$ with $f_G=\ud f/\ud G$, {\it etc.}. In Section \ref{analy}, the properties of the phase space and their cosmological implications are determined. Interestingly, instead of critical points, we obtain lines of critical points in the phase space. We then establish conditions for cosmologically viable $f(G)$ dark energy models in the $m$-$r$ plane. Consequently, when deciding the cosmological viability of a particular $f(G)$ model, one's first resort may be to inspect its $m(r)$ curve in the $m$-$r$ plane.  The particular results of \cite{deftsu08} can be easily recovered from the more general conditions obtained with our approach. In Section \ref{spenum}, we investigate several specific $f(G)$ models and examine their cosmological viability by numerical simulations. Cosmologically viable trajectories of the universe are obtained. We summarise the results in Section \ref{conclu}.


\section{Dynamical Framework for $f(G)$ Cosmology}\label{dframe}

We consider the following action
\be
\label{action1}
S=\int \ud^4 x\sqrt{-g}\left(\frac{1}{2}R+f(G) + {\cal L}_r+{\cal L}_m\right)\,,
\ee
where $R$ is the Ricci scalar, $f(G)$ is a general function of the Gauss-Bonnet term $G$ and $\sqrt{-g}{\cal L}_r$ and $\sqrt{-g}{\cal L}_m$ are the radiation and matter Lagrangian densities respectively. We have chosen $8\pi G_b=1$, where $G_b$ is a bare gravitational constant. Varying the action (\ref{action1}) with respect to $g_{\mu\nu}$, we obtain the field equations for $f(G)$ gravity
\begin{eqnarray}
\label{geeq}
R_{\mu \nu}-\frac{1}{2}R g_{\mu \nu} & &\nn
+8 [ R_{\mu \rho \nu \sigma} +R_{\rho \nu} g_{\sigma \mu}
-R_{\rho \sigma} g_{\nu \mu} -R_{\mu \nu} g_{\sigma \rho}+
R_{\mu \sigma} g_{\nu \rho}& &\nn
+\frac{1}{2} R (g_{\mu \nu} g_{\sigma \rho}
-g_{\mu \sigma} g_{\nu \rho}) ] \nabla^{\rho} \nabla^{\sigma} f_G
+(G f_G-f) g_{\mu \nu}&=&T_{\mu \nu}\,,
\end{eqnarray}
where $T_{\mu\nu}$ is the energy momentum tensor of radiation (energy density $\rho_r$) and matter (energy density $\rho_m$). In the observationally favoured flat Robertson-Walker metric \cite{wmap5}
\be
{\rm d}s^2 = -{\rm d}t^2+a(t)^2 {\rm d}{\bf x}^2\,,
\ee
we have
\be
R=6(\dot H + 2H^2)\,,\quad\quad\quad G=24H^2 (\dot H + H^2)\,,
\ee
and Eqs.~(\ref{geeq}) reduce to
\bea
\label{frieq}
3H^2 &=& G f_G-f-24H^3 \dot{f_G}+\rho_r+\rho_m\,,\\
\label{acceq}
-2\dot{H} &=& -8H^3 \dot{f_G}+16H\dot{H}\dot{f_G}+8H^2 \ddot{f_G}+\frac{4}{3}\rho_r+\rho_m\,,
\eea
where $H$ is the Hubble parameter and an overdot stands for a derivative with respect to $t$. Additionally, the densities $\rho_r$ and $\rho_m$ satisfy the usual continuity equations
\bea
\label{rceq}
\dot{\rho_r} + 4H\rho_r &=& 0\,, \\
\label{mceq}
\dot{\rho_m} + 3H\rho_m &=& 0\,. 
\eea
Eqs.~(\ref{frieq}-\ref{mceq}) determine the dynamics of the $f(G)$ gravity system (\ref{action1}) in a homogeneous and isotropic background.

Drawing the analogy with $3H^2=\rho$ and $-2\dot{H}=p+\rho$, we can naturally define gravitationally induced forms of dark energy density $\rho_{DE}$ and pressure $p_{DE}$ as
\bea
\rho_{DE} &=& G f_G-f-24H^3 \dot{f_G}\,,\\
p_{DE} &=& 16H^3 \dot{f_G}+16H\dot{H}\dot{f_G}+8H^2 \ddot{f_G}-G f_G+f\,.
\eea
In this way, the continuity equation of dark energy holds, 
\be
\dot{\rho_{DE}} + 3H(p_{DE}+\rho_{DE}) = 0\,,
\ee
and its equation of state parameter becomes
\be
w_{DE}=\frac{p_{DE}}{\rho_{DE}}=\frac{16H^3 \dot{f_G}+16H\dot{H}\dot{f_G}+8H^2 \ddot{f_G}-G f_G+f}{G f_G-f-24H^3 \dot{f_G}}\,.
\ee
We also define the effective equation of state \cite{fofr}
\be
w_{eff}=-1-\frac{2\dot{H}}{3H^2}\,,
\ee
and dimensionless energy densities
\be
\Omega_X=\frac{\rho_X}{3H^2}\,,\quad X=r,m,DE\,.
\ee

To discuss the dynamics of a general $f(G)$ model, it is convenient to rewrite the equations of motion as a dynamical system \cite{Halliwell:1986ja,clw,Zhou:2007xp}. To achieve this we introduce the following dimensionless variables
\bea
x_1 &=& \frac{Gf_G}{3H^2}\,,\\
x_2 &=& -\frac{f}{3H^2}\,,\\
x_3 &=& -8H\dot{f_G}\,,\\
x_4 &=& \Omega_r =\frac{\rho_r}{3H^2}\,,\\
x_5 &=& \frac{G}{24H^4}=\frac{\dot{H}}{H^2}+1\,.
\eea
Then Eqs.~(\ref{frieq}-\ref{mceq}) can be recast as a first order dynamical system
\bea
\label{xeq1}
\frac{\ud x_1}{\ud N} &=& -\frac{x_3 x_5}{m}-x_3 x_5-2x_1 x_5 + 2x_1\,,\\
\label{xeq2}
\frac{\ud x_2}{\ud N} &=& \frac{x_3 x_5}{m}-2x_2 x_5+2x_2\,,\\
\label{xeq3}
\frac{\ud x_3}{\ud N} &=& -x_3+2x_5-x_3 x_5+1-3x_1-3x_2+x_4\,,\\
\label{xeq4}
\frac{\ud x_4}{\ud N} &=& -2x_4 - 2x_4 x_5\,,\\
\label{xeq5}
\frac{\ud x_5}{\ud N} &=& -\frac{x_3 x_5^2}{x_1 m}-4x_5^2+4x_5\,,
\eea
where $N=\mathrm{ln}(a/a_i)$ ($a_i$ is the initial value of the scalar factor) and
\bea
\label{mfun}
m &=& \frac{Gf_{GG}}{f_G}\,,\\
\label{rfun}
r &=& -\frac{Gf_G}{f}=\frac{x_1}{x_2}\,.
\eea
Note that Eq.~(\ref{frieq}) is recast as
\be
\Omega_m=1-x_1-x_2-x_3-x_4\,.
\ee
In terms of $x_i$, we can rewrite $w_{DE}$ and $w_{eff}$ as
\bea
w_{DE} &=& \frac{-2x_5-x_4-1}{3(x_1+x_2+x_3)}\,,\\
w_{eff} &=& -\frac{1}{3}(2x_5+1)\,.
\eea

From the r.h.s.~of Eq.~(\ref{xeq4}), we can factor out $x_4$, which suggests that $x_4=0$ is an invariant submanifold \cite{dysy} of the dynamical system, meaning the system can not go through the subspace $x_4=0$ and can only approach it asymptotically. We can also factor out $x_5$ from the r.h.s.~of Eq.~(\ref{xeq5}). However, $x_5=0$ is not an invariant submanifold, because when $x_5=0$ (i.e., $G=0$), the first term ($-x_3 x_5^2/x_1 m=\dot{G}/24H^5$) on the r.h.s.~of Eq.~(\ref{xeq5}) is not necessarily zero.

From Eq.~(\ref{rfun}) we can express $G$ as a function of $x_1/x_2$, and then substituting it into Eq.~(\ref{mfun}), $m$ can be expressed in terms of $x_1/x_2$. Therefore, given a form of $f(G)$, we obtain a function $m(r=x_1/x_2)$ and the dynamical system (\ref{xeq1}-\ref{xeq5}) becomes autonomous. For example, the model $f(G)=\alpha (G^p-\beta)^q$ corresponds to the straight line
$m(r)=(1-q)r/q  + p-1$ in the $m$-$r$ plane. A similar function of $m(r)$ exists in the $f(R)$ models and examining the cosmological viability of $f(R)$ dark energy models according to the corresponding $m(r)$ curve in the $m$-$r$ plane has proven to be a fruitful approach \cite{fofr}. On the other hand, if we know a form of $m(r)$ that is cosmologically viable, one may use it to obtain $f(G)$ from Eqs.~(\ref{mfun}) and (\ref{rfun}).

From Eqs.~(\ref{xeq2}) and (\ref{xeq5}), given a form of $f(G)$, we may express $G$ and $H$ in terms of $x_2$ and $x_5$, and substituting it into Eq.~(\ref{xeq1}), we may then express $x_1$ in terms of $x_2$ and $x_5$, so $x_1$ is actually not an independent variable. However, beyond the power-law form $f(G)=\alpha G^n$, which allows $x_1$ to be eliminated directly, even simple forms of $f(G)$ yield a complicated dependence for $x_1$ on $x_2$ and $x_5$, which is often impossible to obtain in closed form.
Note that in \cite{deftsu08}, the authors recast the field equations into a 4 dimensional dynamical system (autonomous if the form of $f(G)$ is given). However, phase space analysis, especially for general $f(G)$ models, in that reformulation is not easy. Consequently we solve the original 5 dimensional system (\ref{xeq1}-\ref{xeq5}), which is analytically simple and numerically more reliable and efficient.


\section{Analytical Results of General $f(G)$ Models}\label{analy}

Writing the field equations in terms of the new dimensionless variables $x_i$ makes it easy to look at the analytical properties of general $f(G)$ models. The critical points (equilibria) are where the r.h.s.~of the first order dynamical system (\ref{xeq1}-\ref{xeq5}) is zero. A new feature emerges here. Rather than having isolated critical points, we obtain continuous lines of critical points (equilibrium manifolds), which we call \emph{critical lines}:
\bea
L_1 &:& \{x_1=1-x_2,x_2=x_2,x_3=0,x_4=0,x_5=1\}\,,\nn
&\phantom{:}& \Omega_m=0\,,\quad \Omega_r=0\,,\quad\Omega_{DE}=1\,,\quad w_{DE}=-1\,,\quad w_{eff}=-1\,,\nn
L_2 &:& \{x_1=\frac{1}{6}x_3,x_2=-\frac{1}{3}x_3,x_3=x_3,x_4=0,x_5=-\frac{1}{2},m=-\frac{1}{2}\}\,,\nn
&\phantom{:}& \Omega_m=1-\frac{5}{6}x_3\,,\quad \Omega_r=0\,,\quad\Omega_{DE}=\frac{5}{6}x_3\,,\quad w_{DE}=0\,,\quad w_{eff}=0\,,\nn
L_3 &:& \{x_1=\frac{x_5}{x_5-2},x_2=-\frac{2x_5}{x_5-2},x_3=\frac{2(x_5-1)}{x_5-2},x_4=0,x_5=x_5,m=-\frac{1}{2}\}\,,\nn
&\phantom{:}& \Omega_m=0\,,\quad \Omega_r=0\,,\quad\Omega_{DE}=1\,,\quad w_{DE}=-\frac{2}{3}x_5-\frac{1}{3}\,,\quad w_{eff}=-\frac{2}{3}x_5-\frac{1}{3}\,,\nn
L_4 &:& \{x_1=\frac{1}{4}x_3,x_2=-\frac{1}{2}x_3,x_3=x_3,x_4=1-\frac{3}{4}x_3,x_5=-1, m=-\frac{1}{2}\}\,,\nn
&\phantom{:}& \Omega_m=0\,,\quad \Omega_r=1-\frac{3}{4}x_3\,,\quad\Omega_{DE}=\frac{3}{4}x_3\,,\quad w_{DE}=\frac{1}{3}\,,\quad w_{eff}=\frac{1}{3}\,. \nonumber
\eea
$L_1$, $L_2$ and $L_4$ are straight lines in the phase space, while $L_3$ is not. Since $L_1$, $L_2$ and $L_4$ have different constant values of $x_5$, $1$, $-1/2$ and $-1$ respectively, they do not intersect. However, $L_3$ intersects with $L_1$ at $(-1,2,0,0,1)$,  with $L_2$ at $(1/5,-2/5,6/5,0,-1/2)$ and with $L_4$ at $(1/3,-2/3,4/3,0,-1)$.

Generally, if a nonlinear system has a critical line, the Jacobian matrix of the linearised system at a critical point on the line has a zero eigenvalue with an associated eigenvector tangent to the critical line at the chosen point. This kind of nonlinear system is a special subclass of the non-hyperbolic system (whose linearised system has one or more eigenvalues with zero real parts.). The emergence of equilibrium manifolds may be due to some symmetry in the nonlinear system or that the  nonlinear system can be reduced to a lower dimensional one. For the case of $f(G)=\alpha G^n$, as we shall see in Section \ref{spenum}, when the system is reduced to a lower 4 dimensional one with just $x_2$-$x_5$, $L_1$ shrinks to a point but the other 3 lines remain. The stability of a particular critical point on the line can be determined by the non-zero eigenvalues, because near this critical point there is essentially no dynamics along the critical line (i.e., along the direction of the eigenvector associated with the zero eigenvalue), so the dynamics near this critical point may be viewed in a reduced phase space obtained by suppressing the zero eigenvalue direction. It is also interesting to note that the dynamical system (\ref{xeq1}-\ref{xeq5}) does not have any parameters in it, but, as we shall see below, the stability of a critical point changes along the critical line. Thus this system might be considered as the case of \emph{bifurcation without parameters} \cite{bwop}.

Note that for $L_2$, $L_3$ and $L_4$ to exist, $m$ should be equal to $-1/2$, as we have noted in the definitions of the critical lines above. Also, for $L_2$, $L_3$ and $L_4$, we have $r=x_1/x_2=-1/2$. Hence for these critical lines,
\be \label{constraint1}
m(-\frac{1}{2})=-\frac{1}{2}
\ee
is satisfied, i.e., they must be located at the point $(-1/2,-1/2)$ in the $m$-$r$ plane. Thus for $L_2$, $L_3$ and $L_4$ to exist, $r=-1/2$ should solve the equation $m(r)=-1/2$. From the values of $\Omega_{DE}$, $w_{DE}$, {\it etc.}, we can see that $L_2$ and $L_4$ correspond to solutions in which dark energy scales with matter and radiation respectively, thus they can hopefully correspond to the standard matter and radiation epochs. Therefore, for a standard matter or radiation epoch to exist, the constraint (\ref{constraint1}) should be satisfied, i.e., the $m(r)$ curve should pass through $(-1/2,-1/2)$ in the $m$-$r$ plane.

Following the case in general $f(R)$ models \cite{fofr}, we can derive the following equation from Eqs.~(\ref{mfun}) and (\ref{rfun})
\be \label{rr1constraint}
\frac{\ud r}{\ud N}=r(m+r+1)\frac{\dot{G}}{HG}\,.
\ee
When $r=0$ is satisfied and $(m+r+1)\dot{G}/HG$ does not diverge or $m(r)=-r-1$ is satisfied and $r\dot{G}/HG$ does not diverge, we have $\ud r/\ud N=0$. From the $m$-$r$ plane's view, this means that, evolving along the curve $m(r)$, typically the system can not go through any intersection points between the curve $m(r)$ and the $m$ axis ($r=0$) or the particular straight line $m_c(r)=-r-1$. However, the curve $m(r)$ can be ``connected'' at $r=\pm \infty$ or $m=\pm \infty$. So for example, the system may evolve from the $r<0$ half plane to the $r>0$ half plane by ``tunnelling'' through the infinite point of the $m(r)$ curve at $r=\pm \infty$. This happens because basically $r$ is not an essential dynamical variable of the dynamical system (\ref{xeq1}-\ref{xeq5}). For example, $r$ will go through the infinite point if $x_2$ goes from $0^-$ to $0^+$ with $x_1$ remaining finite.

Let us now discuss the properties and cosmological consequences of an arbitrary point on each of the critical lines in turn. Note that supposing the system starts from somewhere near a particular critical line and is initially attracted to the critical line, which particular critical point on the line the system will finally evolve to or pass nearby largely depends on the initial conditions. We denote an arbitrary point on, for example, $L_1$ as $P_1$.
\begin{itemize}
 \item $P_1$: $(1-x_{20},x_{20},0,0,1)$, \emph{de Sitter point}\\
       This is a de Sitter point for any point along the line $L_1$, i.e., for any value of $x_{20}$ (a given value of $x_2$). The eigenvalues of the linearised system are:
       \be
       0\,,\quad -4\,,\quad -3\,,\quad -\frac{3}{2}\pm \frac{1}{2}\sqrt{25+\frac{8}{(x_{20}-1)m_1}}\,,
       \ee   
       where $m_1=m(r_1=(1-x_{20})/x_{20})$. Hence this is a stable spiral in the subspace of the last two eigenvalues when
       \be \label{dsspiral}
       0<m_1<\frac{8}{25(1-x_{20})}\; (x_{20}<1)\quad \mathrm{or} \quad \frac{8}{25(1-x_{20})}<m_1<0 \;(x_{20}>1)\,,
       \ee
       a stable node when
       \be  \label{dsnode}
       \frac{8}{25(1-x_{20})}<m_1<\frac{1}{2(1-x_{20})}\;(x_{20}<1)\phantom{\,,}
       \ee
       or
       \be
       \frac{1}{2(1-x_{20})}<m_1<\frac{8}{25(1-x_{20})}\;(x_{20}>1)\,,
       \ee
       and an unstable point otherwise. In the $m$-$r$ plane, the stable spiral condition (\ref{dsspiral}) corresponds to the area enclosed by the $r$ axis and the curve $m_{dS*}(r)=8(r+1)/25r$, whilst the stable node condition (\ref{dsnode}) corresponds to the area enclosed by the curve $m_{dS*}(r)=8(r+1)/25r$ and the curve $m_{dS}(r)=(r+1)/2r$. Note that the point $(-1/2,-1/2)$ is on the curve $m_{dS}(r)=(r+1)/2r$ (the boundary of the stable area), so the stable area of $L_1$ is connected to $L_2$, $L_3$ and $L_4$ in the $m$-$r$ plane, as seen in Fig.~\ref{mofr2}. One can check that these conditions are consistent with those established for a stable de Sitter point in \cite{deftsu08}. For example, from the condition (\ref{dsspiral}), we can derive the condition for the existence of a stable spiral in \cite{deftsu08}
       \be \label{dssp2}
       0<H_1^6f_{GG}(G_1)<\frac{1}{600}\,,
       \ee
       where $H_1$ and $G_1$ are the Hubble parameter and the Gauss-Bonnet term at the de Sitter point respectively and we have used
       \be
       m_1=\frac{G_1f_{GG}(G_1)}{f_G(G_1)}\,,\quad 1-x_{20}=x_{10}=\frac{G_1f_{G}(G_1)}{3H_1^2}\,,\quad G_1=24H_1^4\,.
       \ee

       \begin{figure}
       \includegraphics[height=5.7in,width=5.7in]{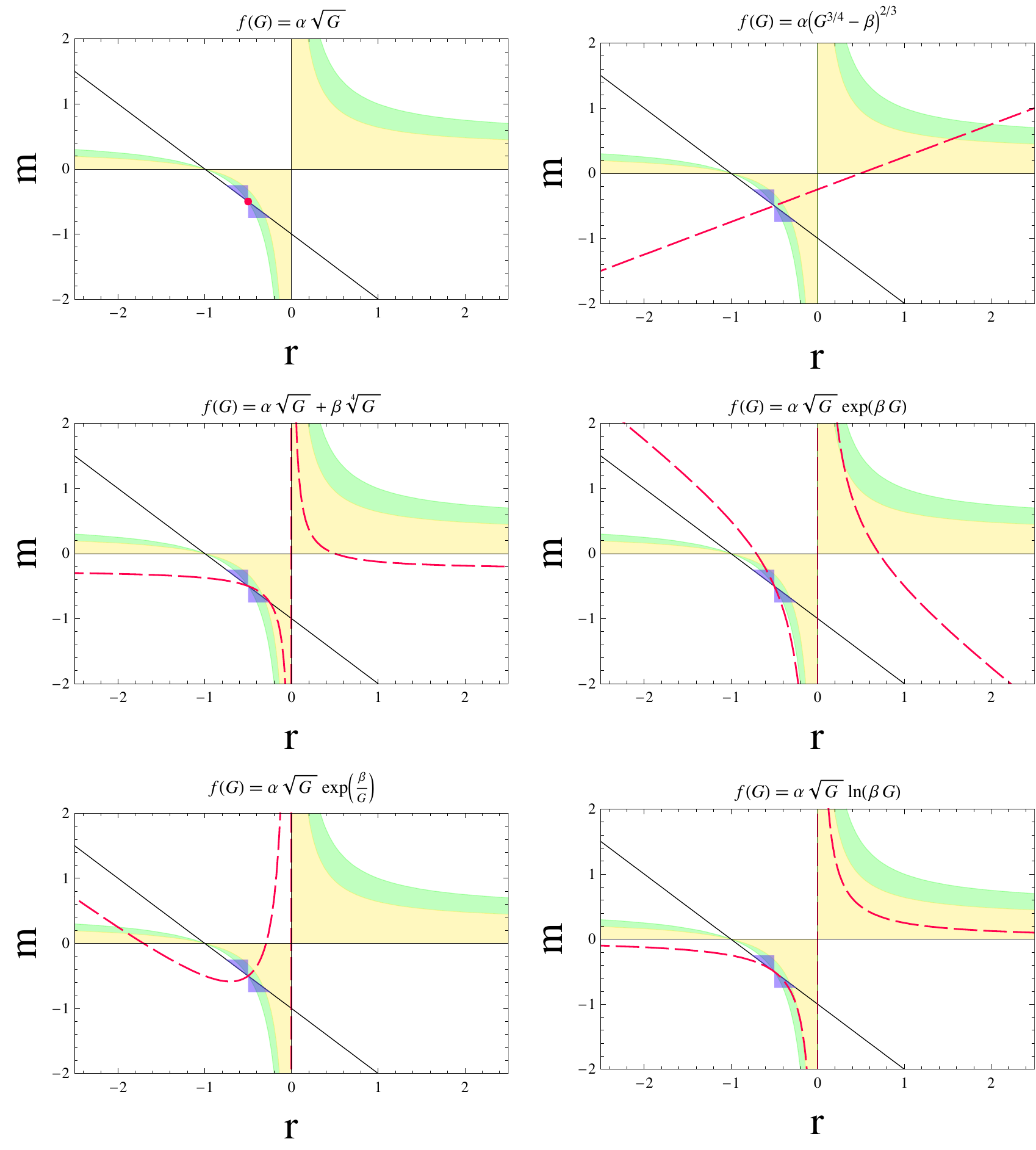}
       \caption{The $m(r)$ curves (red dashed lines) for several $f(G)$ models. Since these $f(G)$ models involve functions such as $\sqrt{G}$, $G^{1/4}$ or $\ln (\beta G)$, $G$ in $f(G)$ should be understood as $|G|$. Note that the matter (radiation) point $P_2$ ($P_4$) and the phantom-like point $P_3$ are located at $(-1/2,-1/2)$. To have a standard matter dominated epoch, the curve should pass through the point $(-1/2,-1/2)$, whilst not accessing the point from the forbidden directions (illustrated by the small light blue area around the point). The light green and light yellow areas are the potentially stable node and stable spiral regions of the de Sitter point $P_1$ respectively. Evolving along the curve $m(r)$, typically the system can not go through any intersection points with the $m$ axis ($r=0$) and the particular straight line $m_c(r)=-r-1$ (black solid line). However, the system can evolve through infinite points of the $m(r)$ curve, as described in the main text.} \label{mofr2}
       \end{figure}

 \item $P_2$: $(\frac{1}{6}x_{30},-\frac{1}{3}x_{30},x_{30},0,-\frac{1}{2})$, \emph{scaling with matter   point}\\
       At this point, dark energy mimics the evolution of matter, and the density of dark energy scales with that of matter ($\Omega_{m}/\Omega_{DE}=(6-5x_{30})/5x_{30}$), whereas there is no radiation. The eigenvalues of the linearised system are:
       \be
       0\,,\quad -1\,,\quad 3(m'(-\frac12)+1)\,,\quad -\frac{3}{4}\pm \frac{1}{4}\sqrt{\frac{96-71x_{30}}{x_{30}}}\,,
       \ee
       where $m'(-1/2)=\ud m(r)/\ud r|_{r=-1/2}$. Along this line, the stability of the last two eigenvalues changes at the points $x_{30}=0$ or $x_{30}=96/71$, which we may call \emph{bifurcation points}: the points with $x_{30}<0$ or $x_{30}>96/71$ are stable, while other points are unstable. In order for $P_2$ to be a standard matter era ($\Omega_m \simeq 1$), $x_{30}$ needs to be near $0$, but $x_{30}=0$ is a singularity of the last two eigenvalues. It is interesting to note that the situation here is quite similar to that of the standard matter point in general $f(R)$ models \cite{fofr}, where for $P_5$ in their notation to be a standard matter era $m_5$ needs to be near $0$, but $0$ is a singularity of two eigenvalues. For our current point $P_2$, if $x_{30}\rightarrow 0^+$, the last two eigenvalues diverge as $-\infty$ and $+\infty$. However, if $x_{30}$ is some small (non-infinitesimal) positive value, the two eigenvalues are large and finite with opposite signs. If the system flows towards these kind of points, unless the initial conditions are extremely fine-tuned, the system will not remain around it for a reasonable long time. Since an elongated matter dominated epoch is needed for cosmic structure formation, we generally consider this case as unacceptable. On the other hand, if $x_{30}\rightarrow 0^-$, then the last two eigenvalues are complex with negative real parts, which means the two eigenvalues are stable. However, the imaginary parts diverge, and hence the system oscillates rapidly in the subspace associated with the two eigenvalues when it approaches the standard matter point. The frequency of the oscillation can be reduced if $x_{30}$ is slightly less than $0$, while the amplitude of the oscillation will not be very large if the initial conditions are set near the critical point in the associated subspace. The system should finally leave the standard matter point and enter an accelerated point, so we shall set $3(m'(-1/2)+1)$ to be greater than $0$ in order to make the matter point unstable, i.e.,
       \be \label{slow2}
       m'(-\frac12)>-1\,.
       \ee
       In the $m$-$r$ plane, it means that there are unacceptable directions of the $m(r)$ curve approaching the point $(-1/2,-1/2)$, see Fig.~\ref{mofr2}. Note that the properties and their cosmological implications obtained above assume that the system is exactly at a critical point. Since the critical point here for the standard matter epoch is not supposed to be totally stable, the system will just pass nearby the critical point rather than fall onto it, so the properties and their cosmological implications when the system passing by the critical point should be just approximate to what are inferred from the critical point analysis.

 \item $P_3$: $(\frac{x_{50}}{x_{50}-2},-\frac{2x_{50}}{x_{50}-2},\frac{2(x_{50}-1)}{x_{50}-2},0,x_{50})$, \emph{dark energy dominated point}\\
       This is a dark energy dominated point with no radiation and no matter. It has a fascinating wealth of possibilities, mimicking radiation when $x_{50}=-1$ and matter when $x_{50}=-1/2$; it is a quintessence-like point ($-1<w_{DE}<-1/3$) if $0<x_{50}<1$ and a phantom-like point ($w_{DE}<-1$) if $x_{50}>1$. The eigenvalues of the linearised system are:
       \be
       0\,,\quad -2x_{50}-1\,,\quad -x_{50}-2\,,\quad -2x_{50}-2\,,\quad -2(x_{50}-1)(m'(-\frac12)+1)\,.
       \ee
       This is a stable point when either of the following two conditions is satisfied
       \be
       -\frac{1}{2}<x_{50}<1\quad \mathrm{and} \quad m'(-\frac12)<-1\,,
       \ee
       or
       \be \label{phantom3}
       x_{50}>1\quad \mathrm{and} \quad m'(-\frac12)>-1\,.
       \ee
       Note that the second condition (\ref{phantom3}) is consistent with the condition (\ref{slow2}), which means that if we require an unstable direction for the system to leave the standard matter point $P_2$, then $P_3$ can be a stable phantom-like point, co-existing with the stable de Sitter point $P_1$. Which point the universe finally evolves to depends on the initial conditions, as we shall see in Section \ref{spenum}. Note that although dark energy with $w<-1$ is problematic theoretically \cite{chm03jcline}, it is still consistent with current data \cite{ckpcb,asss1,wmap5}.

 \item $P_4$: $(\frac{1}{4}x_{30},-\frac{1}{2}x_{30},x_{30},1-\frac{3}{4}x_{30},-1)$, \emph{scaling with radiation point}\\
       At this point, dark energy mimics the evolution of radiation, and the density of dark energy scales with that of radiation ($\Omega_{r}/\Omega_{DE}=(4-3x_{30})/3x_{30}$), whereas there is no matter. The eigenvalues of the linearised system are:
       \be
       0\,,\quad 1\,,\quad 4(m'(-\frac12)+1)\,,\quad -\frac{1}{2}\pm \frac{1}{2}\sqrt{\frac{64-47x_{30}}{x_{30}}}\,.
       \ee 
       The stability structure in the subspace of the last two eigenvalues is similar to that of $P_2$: the points with $x_{30}<0$ or $x_{30}>64/47$ are stable in the subspace (still unstable in the total phase space); in order for $P_4$ to be a standard radiation point ($\Omega_r \simeq 1$), $x_{30}$ needs to be near $0$, but at the same time, $x_{30}=0$ is a singularity of the two eigenvalues; if $x_{30}$ is slightly greater than $0$, $P_2$ is extremely unstable; if $x_{30}$ is slightly less than $0$, $P_2$ is stable in the subspace but the system oscillates significantly when flowing to it. Note that the condition $f_{GG}>0$ from \cite{deftsu08} for the existence of a standard radiation/matter point can also be recovered. For example, to avoid the violent instability of the standard radiation point,  we require $x_{30}$ be slightly less than $0$, so $-f/3H^2=x_{20}=-x_{30}/2>0$, which means $f<0$. Additionally, for the standard radiation point, we have $r=-Gf_G/f=-1/2<0$ and $m(-1/2)=Gf_{GG}/f_G=-1/2<0$, so $f_{GG}>0$ holds at the standard radiation point.
\end{itemize}

Starting from the radiation dominated era, a \emph{cosmologically viable} trajectory of the universe in the phase space would start somewhere near the standard radiation point $P_4$ (with $x_{30}$ slightly less than $0$), then  slowly pass nearby the standard matter point $P_2$ (with $x_{30}$ slightly less than $0$) and finally land at the stable de Sitter point $P_1$ or the stable phantom-like point $P_3$. In the $m$-$r$ plane, with the trajectory fixed along the curve $m(r)$ and subject to the constraint from Eq.~(\ref{rr1constraint}), the universe would start near the point $(-1/2,-1/2)$ where the standard radiation point $P_4$ and the standard matter point $P_2$ are located, slowly moving away from it, and finally fall onto some point in the stable area of $P_1$ or come back to evolve to $P_3$ that also resides at the point $(-1/2,-1/2)$. Note that it might also be possible that the universe has an unstable accelelated epoch after the matter dominated period, whose cosmological viability is not the concern of this paper and is left for future work. Indeed, for phase spaces higher than 2 dimensions, more complicated dynamical phenomena such as chaos may appear, but again this is beyond the scope of the current paper.


\section{Specific Models and Numerical Results}\label{spenum}

Having described possible cosmic trajectories for general $f(G)$ dark energy models, we now turn our attention to the cosmological viability of a few specific toy models of $f(G)$ whose $m(r)$ curves can be analytically obtained. Note that if the analytical form of $m(r)$ can not be obtained, one may get a numerical expression of it for a certain range of $r$. On the other hand, given a form of $m(r)$, it may not be easy to obtain an analytical form for $f(G)$, but again we can solve it numerically.

The Gauss-Bonnet term $G$ evolves from negative values in the radiation and matter dominated epochs to positive values in the acceleration epoch, thus the lagrangian is not well defined if there is any function of the Gauss-Bonnet term such as $\sqrt{G}$ or $\ln (\beta G)$. In such cases, $G$ in the $f(G)$ should be understood as $|G|$ in this section as well as in Fig.~\ref{mofr2}. We will start the system deep in the radiation dominated epoch, which requires we initially set $r=x_1/x_2\simeq -1/2$, $x_4\simeq 1$, $x_5\simeq -1$ and $x_1+x_2+x_3\simeq 0$. Since the matter dominated point is located at $(-1/2,-1/2)$, setting $r=x_1/x_2\simeq -1/2$ is also a requirement for the universe to have an elongated matter dominated period. We find that the dynamical systems of $f(G)$ models are quite often stiff. The simulation figures in this section are produced using the Matlab stiff ordinary differential equation solver \emph{ode15s}.

We first consider the simplest case $f(G)=\alpha G^n$. This model gives a point
\be
(r,m)=(-n,n-1)
\ee
rather than a curve in the $m$-$r$ plane, see Fig.~\ref{mofr2}. Since $r=x_1/x_2=-n$, we have $x_1=-nx_2$. So we may consider this system in the 4 dimensional subspace $x_2$-$x_5$. In this subspace, while $L_2,L_3,L_4$ remain critical lines, $L_1$ shrinks to a critical point $(1/(1\!-\!n),0,0,1)$. In order to have the standard matter and radiation points, the point $(-n,n-1)$ should be located at $(-1/2,-1/2)$, thus $n=1/2$. Note that this result is consistent with what has been obtained in surveying the scaling solutions of $f(G)$ models \cite{lidsey08}. So in this case, all the 4 critical manifolds (points or lines) are at $(-1/2,-1/2)$. Re-calculating the eigenvalues, we find that, for $P_1$, they are $-4,-3,0,-3$; for $P_2$, $P_3$ and $P_4$, the eigenvalues containing $m'(-1/2)$ disappear, compared to the 5 dimensional case. The $0$ eigenvalue for $P_1$ here is not related to any critical line, thus the stability of this isolated non-hyperbolic point may be obtained by the centre manifold method \cite{dysy}. Since the eigenvalue containing $m'(-1/2)$ for $P_2$ has vanished, this point is either totally stable (the system will stay in the matter dominated era for ever and not evolve to the dark energy dominated era.) or extremely unstable (the system can not stay in the matter dominated era for long enough to allow cosmic structure to form.). Therefore this model is ruled out unless we allow extreme fine-tuning of initial conditions.

The power-law model may be generalised to $f(G)=\alpha (G^p-\beta)^q$. The corresponding curve $m(r)$ and its first derivative $m'(r)$ for this model are respectively
\be
m(r)=\frac{1-q}q r + p-1\quad {\rm and} \quad m'(r)=\frac{1-q}q\,.
\ee
To have a standard matter point, i.e., to employ the relation (\ref{constraint1}), we find it leads to a constraint equation for the parameters
\be
2pq=1\,.
\ee
The system needs to be able to leave the matter point, which leads to a further constraint
\be
q>0\,.
\ee
Thus if these constraints are satisfied, cosmologically viable trajectories can in principle be obtained.

\begin{figure}
\includegraphics[height=2.9in,width=6.1in]{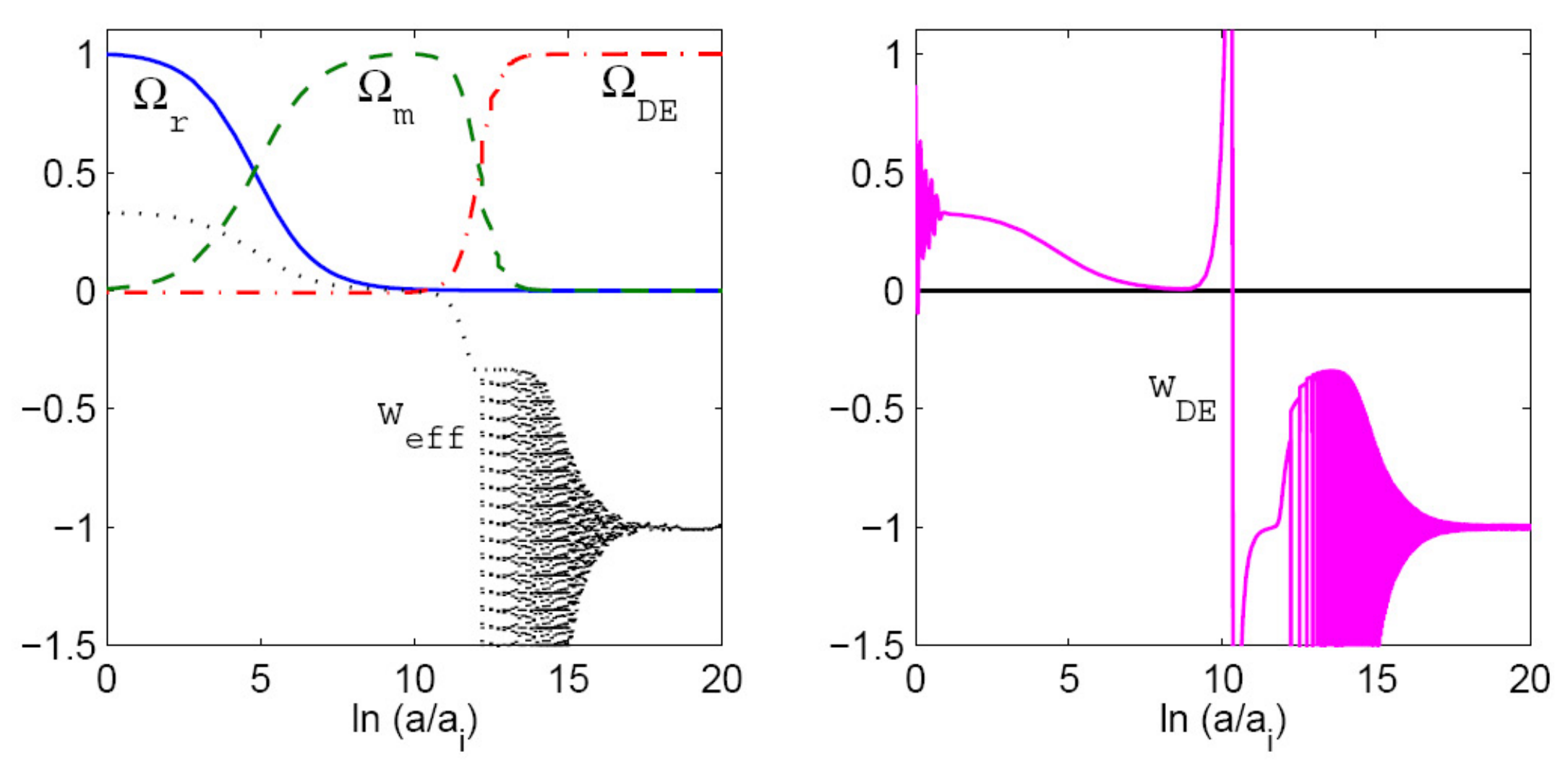}
\caption{The cosmic evolution for the case $f(G)=\alpha (G^\frac{3}{4}-\beta)^\frac{2}{3}$ with initial conditions $x_1=-0.0025 \times (1-10^{-17})$, $x_2=0.005$, $x_3=-0.01$, $x_4=0.99951$ and $x_5=-0.99$. $a_i$ is the initial value of the scalar factor. Note that we have to set $r$ very close to $-1/2$ to obtain a long enough matter dominated period. When oscillating into the de Sitter period, $w_{DE}$ as well as $w_{eff}$ go well below the $w=-1$ divide.} \label{glambda}
\end{figure}

\begin{figure}
\includegraphics[height=2.9in,width=6.1in]{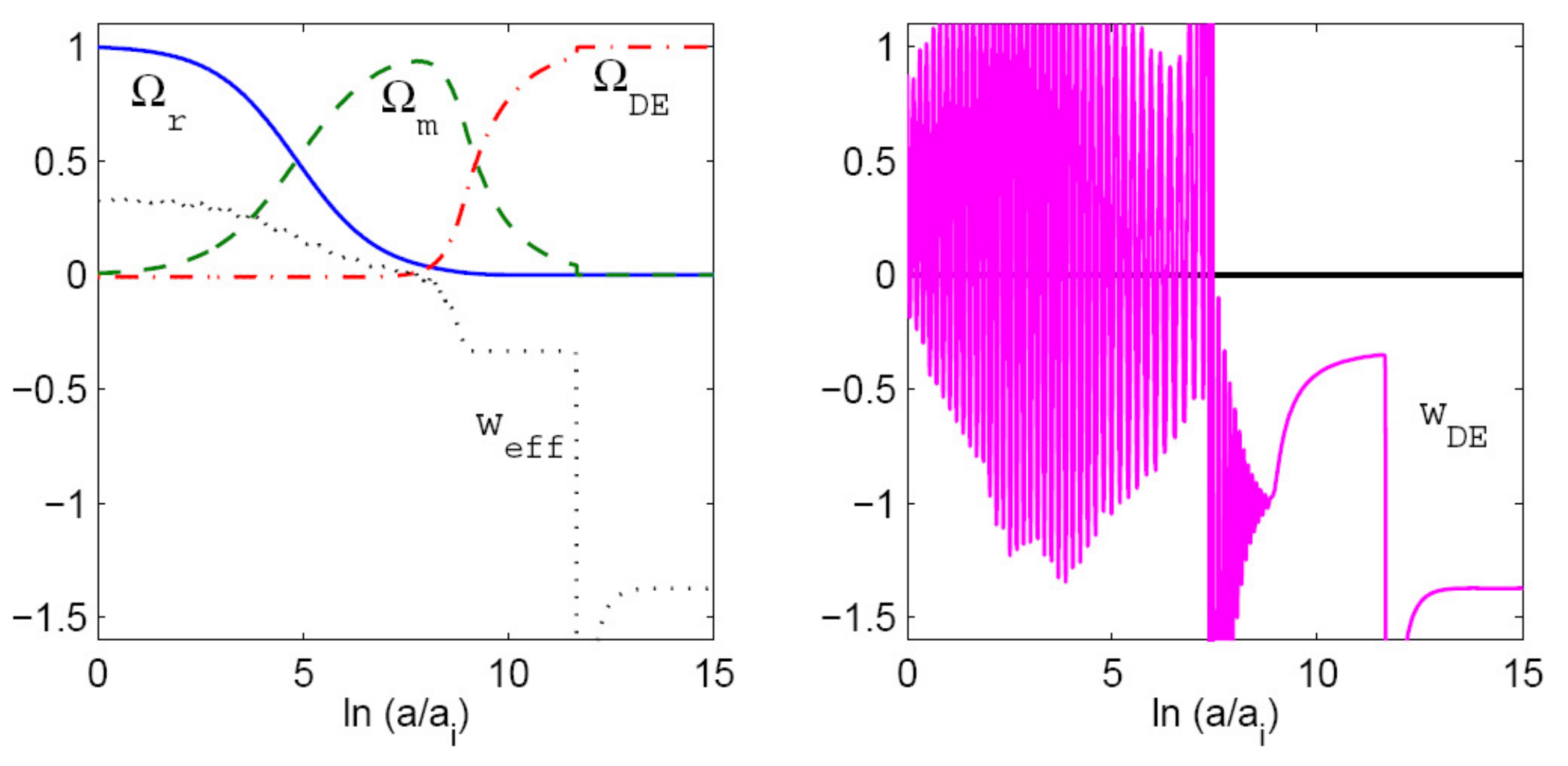}
\caption{The cosmic evolution for the case $f(G)=\alpha (G^\frac{3}{4}-\beta)^\frac{2}{3}$ with initial conditions $x_1=-0.0025 \times (1-5 \cdot 10^{-17})$, $x_2=0.005$, $x_3=-0.0099$, $x_4=0.99951$ and $x_5=-0.99$.} \label{glambda1}
\end{figure}

To be specific, we set $p=3/4$ and $q=2/3$ (see Fig.~\ref{mofr2}) and numerically solve this model to get cosmic trajectories. In Fig.~\ref{glambda}, we obtain a trajectory that is ended with a stable de Sitter era. For this trajectory, while the behaviours of $\Omega_r$, $\Omega_m$, $\Omega_{DE}$ and $w_{eff}$ are quite similar to those of the $\Lambda$CDM model in the radiation and matter dominated epochs, $w_{eff}$ oscillates rapidly and goes deep into the phantom-like region as the universe enters the de Sitter period. On the other hand, $w_{DE}$ at first mimics the background fluids (although oscillating significantly when leaving the radiation period), then at the end of the matter dominated epoch plunging below the $w=-1$ divide for a while, thereafter increasing above the divide, and finally oscillating rapidly into the de Sitter period. In the $m$-$r$ plane, confined on the straight line $m=1/2\:\cdot\:r-1/4$, the universe slowly leaves the point $(-1/2,-1/2)$ towards the $r=0$ line, bounces back and forth and then goes into the stable de Sitter point with $r<0$. The reason why $w_{DE}$ as well as $w_{eff}$ oscillate rapidly when the system enters the de Sitter period is related to the fact that the line $r=0$ actually corresponds to the bifurcation point of $L_1$, similar to the bifurcation point of $L_2$ that is discussed in details in Section \ref{analy}. To see this, we express the last two eigenvalues of $P_1$ in terms of $r$,
\be
-\frac{3}{2}\pm \frac{1}{2}\sqrt{\frac{50r^2-57r-32}{r(2r-1)}}\,,
\ee
and we find that these eigenvalues have divergent imaginary parts if $r \to 0^-$.

Note that the system may also evolve to the stable phantom-like point $P_3$ with $x_{50}>1$ after the matter dominated period, see Fig.~\ref{glambda1}. In this case, $w_{DE}$ oscillates rapidly in the radiation and matter dominated epochs, then plunges well below the $w=-1$ divide at the end of the matter dominated epoch and thereafter shortly passes nearby the de Sitter point $P_1$ (with rapid oscillation) and the point $P_3$ with $x_{50}\simeq 0$ before going into the stable phantom-like point. In the $m$-$r$ plane's view, the system slowly leaves the point $(-1/2,-1/2)$ towards the $r=0$ line and then bounces back and falls onto the point $(-1/2,-1/2)$.

We have also surveyed several other cases. The forms of these models, the corresponding $m(r)$ (see Fig.~\ref{mofr2}) and the constraint of the parameters from Eq.~(\ref{constraint1}) are listed below:
\bea
f(G) = \alpha G^p + \beta G^q: &  m(r)=p +q-1 + pq/r\,,&  p=1/2\,,\nn
f(G) = \alpha G^p\, {\rm exp}(\beta G):  & m(r)=-r + p/r\,,&  p=1/2 \,,\nn
f(G) = \alpha G^p \,{\rm exp}(\beta /G):  & m(r)=-r-p/r-2\,, &  p=1/2 \,,\nn
f(G) = \alpha G^p\,[{\rm ln}(\beta G)]^q: &  m(r)=A(r)/qr\,,& p=1/2\,,\nonumber
\eea
where $A(r)=(1-q)r^2+(2p-q)r+p^2$. The model $f(G)=\alpha G^p\, {\rm exp}(\beta G)$ is ruled out (if extreme fine-tuning of initial conditions is not allowed) because the condition (\ref{slow2}) can not be satisfied, i.e., the corresponding $m(r)$ curve approaches the point $(-1/2, -1/2)$ from a forbidden direction. For the other models, it is easy to produce an elongated matter dominated epoch following a radiation dominated epoch, but we find it difficult for the cosmic trajectory to evolve to a final stable accelerated epoch after the standard matter dominated epoch.


\section{Conclusions}\label{conclu}

In this paper we have derived conditions for cosmologically viable $f(G)$ models. By recasting the field equations of $f(G)$ gravity with a flat Robertson-Walker metric into a 5 dimensional autonomous system, we have seen that the equilibria of the system are not isolated but form so-called equilibrium manifolds (in the present case they are $1$ dimensional lines, called critical lines.). We have shown that the emergence of the critical lines is not totally due to the fact that the 5 dimensional autonomous system can in principle be reduced to a 4 dimensional one. Discovering $4$ critical lines in the phase space for a general $f(G)$ model, the curved line $L_3$ is found to connect to the other 3 straight lines, $L_1$, $L_2$ and $L_4$. We have also found that for all $4$ lines the stability of a critical point changes when moving along the critical line, which may be referred to as bifurcation without parameters. For the critical line $L_2$ (or $L_4$) in which the dark energy density scales with that of matter (or radiation), matter (or radiation) becomes dominant only if the parameter of the critical line $x_{30}$ is around $0$. At the same time, the stability of the critical line changes rapidly around $x_{30}=0$, and we require $x_{30}<0$ to avoid a violent instability (i.e., to avoid extreme fine-tuning of the initial conditions).

Similar to the case of $f(R)$, for a specific $f(G)$ model, we can construct the $m(r)$ curve in the $m$-$r$ plane. In this plane, $L_2$, $L_3$ and $L_4$ all lie at the point $(-1/2,-1/2)$. Thus for a standard matter point $P_2$ (on $L_2$) to exist, we require the $m(r)$ curve satisfy the relation (\ref{constraint1}):
\begin{equation*}
m(-\frac12)=-\frac12\,.
\end{equation*}
The condition (\ref{slow2}):
\begin{equation*}
m'(-\frac12)>-1
\end{equation*}
is also employed to let the universe finally leave the matter dominated point, which corresponds to some forbidden directions around the point $(-1/2,-1/2)$ in the $m$-$r$ plane.

The stable condition for the de Sitter point $P_1$ (on $L_1$) corresponds to the area enclosed by the curve $m_{dS}(r)=(r+1)/2r$ and the $r$ axis in the $m$-$r$ plane, in which the area enclosed by the curve $m_{dS*}(r)=8(r+1)/25r$ and the $r$ axis corresponds to a stable spiral (in a 2 dimensional subspace) with the rest corresponding to a stable node. $P_3$ (on $L_3$) is a dark energy dominated point. It has a wealth of possibilities. Besides mimicking radiation when $x_{50}=-1$ and matter when $x_{50}=-1/2$, it can be a stable quintessence-like point or a stable phantom-like point. But if we require a standard matter epoch (requiring the condition (\ref{slow2}) to hold unless we allow extreme fine-tuning of initial conditions), it can only be stable as a phantom-like point. Which stable accelerated point the universe finally evolves to, the de Sitter point $P_1$ or the phantom-like point $P_3$, depends on the initial conditions.

We have also studied several toy models of $f(G)$ dark energy whose $m(r)$ curves can be obtained analytically. Note that if the analytical form of $m(r)$ can not be obtained, one may obtain it numerically. On the other hand, if a form of $m(r)$ can be obtained from observations or other kinds of analyses, it may not be easy to derive the corresponding $f(G)$, but still we might find it numerically. For the model $f(G)=\alpha (G^p-\beta)^q$ in particular, cosmologically viable trajectories have been obtained for the case with a de Sitter epoch as the final stage as well as for the case with a phantom-like epoch as the final stage. These trajectories mimic the $\Lambda$CDM scenario in the radiation and matter dominated periods (although $w_{DE}$ may oscillate for a certain period) but have distinctive signatures at late times. Most significantly, $w_{DE}$ plunges below the $w=-1$ divide in the late matter dominated epoch (pole-like behaviour, for similar behaviours of $w_{DE}$ in other scenarios see \cite{wpole}) and then oscillates rapidly, also going below the $w=-1$ divide, when evolving to (for the case with a de Sitter epoch as the final stage) or passing nearby (for the case with a phantom-like epoch as the final stage) the de Sitter point $P_1$. For other potentially viable models, cosmologically viable trajectories may still exist, but we find it is non-trivial to obtain them. We leave this task for future work.

\begin{acknowledgments}

We thank Paul Matthews for helpful discusions on dynamical systems with equilibrium manifolds and bifurcation without parameters. We would also like to thank Antonio Padilla for helpful discusions. EJC acknowledges support from the Royal Society. SYZ is supported by scholarships from the University of Nottingham.

\end{acknowledgments}


\end{document}